 \documentclass[twocolumn]{aastex6}
\shorttitle{Black hole wind in the galaxy merger IRAS~F11119$+$3257}
\shortauthors{Tombesi et al.}

\begin{document}



\title{NuSTAR view of the black hole wind in the galaxy merger IRAS~F11119$+$3257}


\author{F. Tombesi$^{1,2,3}$, S. Veilleux$^{2,4}$, M. Mel{\'e}ndez$^{2, 3,
  5}$, A. Lohfink$^{6,7}$, J.~N. Reeves$^8$, E. Piconcelli$^9$,
F. Fiore$^9$, C. Feruglio$^{10,11}$}
\affil{$^1$Department of Physics, University of Rome ``Tor Vergata'', Via della Ricerca Scientifica 1, I-00133 Rome, Italy; francesco.tombesi@roma2.infn.it}
\affil{$^2$Department of Astronomy, University of Maryland, College Park, MD 20742, USA; ftombesi@astro.umd.edu}
\affil{$^3$NASA/Goddard Space Flight Center, Code 662, Greenbelt, MD
  20771, USA; francesco.tombesi@nasa.gov}
\affil{$^4$Joint Space-Science Institute, University of Maryland,
  College Park, MD 20742, USA}
\affil{$^5$Wyle Science, Technology and Engineering Group, 1290
  Hercules Avenue, Houston, TX 77058 USA}
\affil{$^6$Institute of Astronomy, Madingley Road, Cambridge, CB3 0HA,
  UK}
\affil{$^7$Department of Physics, Montana State University, P.O. Box 173840,
Bozeman, MT 59717-3840, USA}
\affil{$^8$Astrophysics Group, School of Physical and Geographical Sciences, Keele University, Keele, Staffordshire, ST5 5BG, UK}
\affil{$^9$INAF Osservatorio Astronomico di Roma, Via Frascati 33, 00078 Monteporzio Catone, Italy}
\affil{$^{10}$INAF Osservatorio Astronomico di Trieste, Via G.B. Tiepolo 11, I-34143 Trieste, Italy}
\affil{$^{11}$Scuola Normale Superiore, Piazza dei Cavalieri 7, I-56126 Pisa, Italy}



\begin{abstract}


Galactic winds driven by active galactic nuclei (AGN) have been invoked to
play a fundamental role in the co-evolution between supermassive black
holes and their host galaxies. Finding observational evidence of such
feedback mechanisms is of crucial importance and it requires a
multi-wavelength approach in order to compare winds at different
scales and phases. In Tombesi et al.~(2015) we reported the detection
of a powerful ultra-fast outflow (UFO) in the \emph{Suzaku} X-ray
spectrum of the ultra-luminous infrared galaxy IRAS~F11119$+$3257. 
The comparison with a galaxy-scale OH molecular outflow observed with
\emph{Herschel} in the same source supported the energy-conserving
scenario for AGN feedback. The main objective of this work is to perform an independent
check of the \emph{Suzaku} results using the higher
sensitivity and wider X-ray continuum coverage of \emph{NuSTAR}. We clearly
detect a highly ionized Fe K UFO in the 100~ks \emph{NuSTAR}
spectrum with parameters $N_H = (3.2\pm1.5)\times 10^{24}$~cm$^{-2}$,
log$\xi$$=$$4.0^{+1.2}_{-0.3}$~erg~s$^{-1}$~cm, and $v_{\text{out}} =
0.253^{+0.061}_{-0.118}$~c. The launching radius is likely at a
distance of $r \ge 16$~$r_s$ from the black hole. The mass outflow rate is in the range
$\dot{M}_{out}$$\simeq$0.5--2~$M_{\odot}$~yr$^{-1}$. The UFO momentum
rate and power are $\dot{P}_{out} \simeq$0.5--2 $L_{AGN}/c$ and
$\dot{E}_{out} \simeq$7--27\% $L_{AGN}$, respectively. The UFO
parameters are consistent between the 2013 \emph{Suzaku} and the 2015
\emph{NuSTAR} observations. Only the column density is found to be
variable, possibly suggesting a clumpy wind. The comparison with
the energetics of molecular outflows estimated in infrared and millimeter wavelengths support a connection between the nuclear and galaxy-scale winds in luminous AGN.

\end{abstract}


\section[]{Introduction}

It has been suggested that mergers between gas-rich galaxies can possibly trigger major
starbursts, enhance the growth of supermassive black holes and,
ultimately, lead to the formation of red gas-poor elliptical galaxies
(e.g. Hopkins et al. 2006). In this galaxy merger scenario, as dust
and gas is gradually dispersed, a completely obscured ultra-luminous
infrared galaxy (ULIRG) evolves to a dusty quasar, and finally to a
completely exposed quasar. Galactic winds driven by the central active
galactic nucleus (AGN) and/or the surrounding starburst have been
invoked to play a fundamental role in this phase, quenching the growth
of both the supermassive black hole (SMBH) and stellar spheroidal
component, and possibly explaining the tight SMBH-spheroid mass
relations (e.g. Silk \& Rees 1998; King 2003; King \& Pounds 2003;
Gaspari \& S{\c a}dowski 2017). Finding observational evidence of
such feedback mechanism(s) is of crucial importance to understand
galaxy and SMBH evolution (e.g., Fiore et al.~2017).

In order to inhibit star formation, outflows have to affect the
molecular gas out of which stars form. Far-infrared molecular
spectroscopy with Herschel of ULIRGs provided a breakthrough in
identifying and analyzing these massive molecular outflows. In
particular, OH absorption observations have revealed molecular
outflows at hundreds of parsec scales with high velocities, up to maximum values of
1,200 km~s$^{-1}$, implying significant mass outflow rates up to
150--1,500 $M_{\odot}$~yr$^{-1}$ (e.g., Sturm et al.~2011; Veilleux et
al.~2013; Gonz{\'a}lez-Alfonso et al.~2014, 2017). 

Several models have been suggested in order to explain their origin,
but essentially all of them require an initial very fast
($v_{\text{out}}$$\sim$0.1c) AGN accretion disk wind driving a shock into the
host galaxy interstellar medium, and the molecular material is removed
by the resultant hot shocked bubble (e.g., Zubovas \& King 2012;
Faucher-Gigu{\`e}re \& Quataert 2012; Costa et al.~2014).
The shock caused by the interaction between the fast wind and the host
galaxy interstellar medium divides the resultant large-scale outflow
in two regimes. Momentum-conserving flows occur if most of the wind
kinetic energy is radiated away. In this case radiation and ram
pressure exert work on the
interstellar medium gas. Instead, energy-conserving flows occur if the
shocked wind gas is not efficiently cooled, and instead expands
adiabatically as a hot bubble. In particular, it is expected that the
momentum flux of the outflow in the energy-conserving case would be
larger than that of the radiation, up to values of $\dot{P} \simeq 10 L_{\text{AGN}}/c$ as
observed in several ULIRGs (e.g., Sturm et al.~2011; Cicone et
al.~2014; Gonz{\'a}lez-Alfonso et al.~2017). However, despite these significant developments,
observational evidences for the connection between the putative nuclear fast AGN
wind and the large-scale outflows are difficult to find.

Blueshifted Fe XXV/XXVI absorption lines at E$>$7~keV have been detected
in the X-ray spectra of several AGNs. Often, the implied velocity is
very high, up to the mildly relativistic values of $v_{\text{out}}$$\simeq$0.1~c for the
so-called ultra-fast outflows (UFOs) (e.g., Tombesi et al. 2010, 2011,
2014, 2015; Gofford et al.~2013; Nardini et al.~2015; Longinotti et
al.~2015; Parker et al.~2017). These winds are observable at sub-parsec scales from the
central black hole, consistently with an accretion disk interpretation, and they
seem to be powerful enough to have a substantial effect on the host
galaxy environment (e.g., Tombesi et al.~2012, 2015; Wagner et al.~2013;
Gofford et al.~2015; Nardini et al.~2015). 

Recently, a 250~ks \emph{Suzaku}
observation of IRAS~F11119+3257 obtained by our group in 2013 showed 
the presence of a powerful UFO with a velocity of
$v_{\text{out}}$$\simeq$0.25c in this ultra-luminous infrared galaxy hosting a large-scale
molecular outflow (Tombesi et al.~2015). Comparing the energetics of these two winds, we
found support for the validity of the energy-conserving mechanism connecting the
inner black hole winds to the large-scale molecular outflows (e.g.,
Zubovas \& King 2012; Faucher-Gigu{\`e}re \& Quataert 2012). An analogous
conclusion was reached by Feruglio et al.~(2015) for the ULIRG/quasar
Mrk~231.

\section[]{The galaxy IRAS~F11119$+$3257}

IRAS~F11119+3257 is a ULIRG at z$=$0.189 with an AGN dominated
quasar-like luminosity of $L_{\text{AGN}} \simeq 1.5\times10^{46}$
erg~s$^{-1}$ (Veilleux et al.~2013). This source was previously observed with \emph{Chandra} for a 15ks snapshot exposure
as part of the quasar and ULIRG evolution study (QUEST) campaign (Teng
\& Veilleux 2010). It provides a very good combination of X-ray brightness and moderate neutral absorption in
the QUEST survey, making it a promising candidate to study the Fe K
band absorbers in this type of objects.

The data to model ratios of our 250~ks \emph{Suzaku} observation obtained in May 2013 with respect to an absorbed ($N_H \simeq
2\times10^{22}$ cm$^{-2}$) power-law continuum ($\Gamma \simeq 2$)
showed the presence of strong absorption residuals in the energy range
$E\simeq$8–-10~keV and a factor of $\sim$3 hard excess at energies of
$E\simeq$15–-25 keV (Tombesi et al.~2015). The absorption is well
modeled with a broad inverted Gaussian line with parameters $E = 9.82^{+0.64}_{-0.34}$ keV,
$\sigma_E = 1.67^{+1.00}_{-0.44}$ keV and equivalent width $EW =
-1.31^{+0.40}_{-0.31}$ keV with a high detection confidence level of
6.5$\sigma$.

If identified with absorption from Fe~XXV--XXVI, this feature would
indicate a high blueshifted velocity of $v_{\text{out}}$$\simeq$0.2–-0.3~c. Indeed,
a fit using a dedicated \emph{XSTAR} table (Kallman \& Bautista 2001) with a velocity broadening of 30,000 km~s$^{-1}$
provides the best representation of the data, compared to an absorption edge or relativistic reflection
(Tombesi et al.~2015). The highly ionized absorber has a colum density
$N_H = (6.4^{+0.8}_{-1.3})\times 10^{24}$ cm$^{-2}$, an ionization
parameter log$\xi = 4.11^{+0.09}_{-0.04}$ erg~s$^{-1}$~cm, and an
outflow velocity of $v_{\text{out}} = 0.255 \pm 0.011$c.

This fast wind model allows us to simultaneously fit both the Fe K
absorption and the hard excess. However, we note that these two
spectral features, fundamental for the interpretation as a fast wind,
may have some ambiguities in the \emph{Suzaku} data given that they
are detected in two separate instruments. In particular, the broad
absorption is at the high energy end of the XIS bandpass and  the
high energy detector PIN is a non-imaging background dominated
instrument. Moreover, the XIS and PIN do not overlap in energy, which
is important for placing the right continuum level at the energies
between E$\simeq$10~keV. An analysis of the \emph{NuSTAR} spectrum would allow to
have an important independent check on the presence of the UFO in this ULIRG and
to simultaneosly constrain the hard X-ray continuum up to energies of E$\simeq$30~keV. 

\section[]{NuSTAR data analysis and results}

Here, we describe the spectral analysis of the \emph{NuSTAR}
observation of IRAS~F11119$+$3257 performed in May 2015 (obsID 60101045002).
The \emph{NuSTAR} data were reduced following the standard procedure, using the
\emph{nupipeline} and \emph{nuproducts} scripts available in the NuSTARDAS software
package. We used NuSTARDAS version 1.4.1 and the calibration
version. We extracted the source spectrum from a circular region with
a 70 arcsec radius, centered on the peak of the point-source image. The
background spectra for each focal plane module (FPM), FPMA and FPMB,
were extracted from a polygonal region on the detector containing the
source and avoiding the area 90 arcsec around the peak. We use the
data taken in standard SCIENCE mode, with a total exposure of 104~ks.

The spectral analysis was performed in the energy interval
E$=$2--20~keV using the software \emph{XSPEC} v.12.8.2. We grouped the
data to a minimum of 25 counts per bin in order to apply the
$\chi^2$-statistic. We performed joint fits for the two
separated FPMA and FPMB detectors and the spectra were merged only for
plotting purposes. In all the models, we include a cross-normalization parameter for the
two detectors, which was found to be within 8\%.  All parameters are given in
the source rest frame and the errors are at the 1$\sigma$ level if not
otherwise stated. Standard Solar abundances are assumed (Asplund et al.~2009). 

In Fig.~1 we show the source spectra of the two detectors and the relative
backgrounds. We note that the source is clearly detected over the
background in the whole E$=$2--20~keV observed energy band and the
spectra taken by the two detectors agree very well in both slope and
normalization. The observed flux is the E$=$2--10~keV and
E$=$10--20~keV  energy bands is $1\times
10^{-12}$~erg~s$^{-1}$~cm$^{-2}$ and $4\times
10^{-13}$~erg~s$^{-1}$~cm$^{-2}$, respectively.

  \begin{figure}[t!]
  \centering
   \includegraphics[width=8.5cm,height=7cm,angle=0]{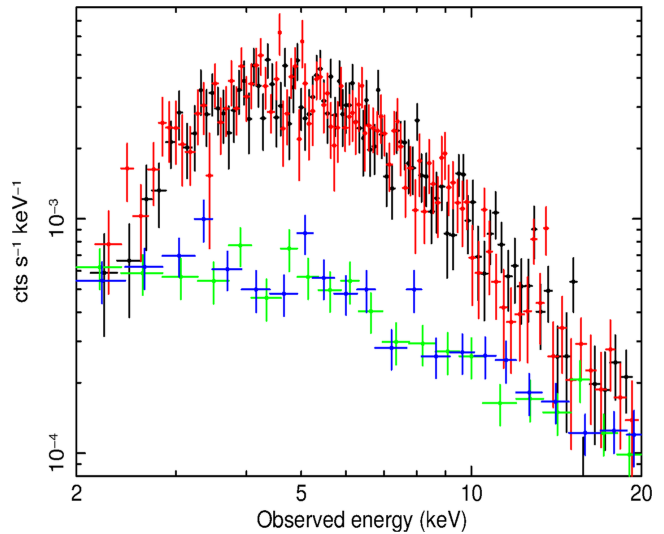}
   \caption{Separated \emph{NuSTAR} FPMA/FPMB source (black/red) and
     background (green/blue) spectra of IRAS~F11119$+$3257. The data are
     rebinned to a minimum of 25 counts per bin and no
     cross-normalization correction is applied.}
    \end{figure}

\subsection[]{Spectral analysis}

A neutral Galactic absorption of $N_H = 2.1\times 10^{21}$~cm$^{-2}$ modeled with \emph{wabs} in
\emph{XSPEC} is included in all the fits (Kalberla et al.~2005). 
We started the spectral modeling with an absorbed continuum power-law with
$\Gamma \simeq 2$ ($\chi^2/\nu = 210/183$). 
The inclusion of an intrinsic neutral absorption component
  \emph{zwabs} of $N_H =
2.1\times 10^{22}$~cm$^{-2}$ as found in \emph{Suzaku} is not
statistically required here because
\emph{NuSTAR} is not sensitive at energies below E$=$2~keV.

In Fig.~2 we show the ratio
of the \emph{NuSTAR} data with respect to the absorbed $\Gamma = 2$
power-law continuum. We can observe broad emission residuals red-ward
of the main Fe K emission lines (vertical lines indicating Fe K$\alpha$ at E$=$6.4~keV, Fe~XXV
He$\alpha$ at E$=$6.7~keV, and the Fe~XXVI Ly$\alpha$ at E$=$6.97~keV,
respectively). We also note the presence of absorption residuals
blue-ward of the main Fe K lines, likely associated with blueshifted
highly ionized Fe resonance lines and edges. The Fe K emission and
absorption are reminiscent of a P-Cygni line profile from a wind, as
recently reported by Nartini et al.~(2015) for the combined
\emph{XMM-Newton} and \emph{NuSTAR} spectrum of the quasar PDS~456.

\subsection[]{Accretion disk wind}

The main objective of this analysis is to check if the \emph{NuSTAR}
data independently confirm the wind model found to provide the
best fit for the \emph{Suzaku} observation of
IRAS~F11119$+$3257 performed in 2013 (Tombesi et al.~2015). 

In order to model the Fe K absorption we then included a dedicated
photoionization table using the \emph{XSTAR} code version
2.2.1bn (Kallman \& Bautista 2001). We consider a $\Gamma = 2$ power-law continuum, consistent
with the observed value, and standard Solar abundances (Asplund et al.~2009). We tested
different broadenings due to turbulent velocities of
5,000~km~s$^{-1}$, 10,000~km~s$^{-1}$, and 30,000~km~s$^{-1}$. The
latter provides the best fit and it is consistent with the large
width observed for the absorption feature. This table is equivalent to the one used by Tombesi et al.~(2015) to
model the \emph{Suzaku} spectrum. 

This high turbulence value is introduced only to model the width of
the absorption line and it is probably not linked to an actual
physical turbulence in the gas. For instance, detailed accretion-disk wind models show that the line profile becomes much
broader because of the velocity shear between consecutive zones of
the wind (e.g., Fukumura et al.~2010, 2014, 2015, 2017).

  \begin{figure}[t!]
  \centering
   \includegraphics[width=8.5cm,height=7cm,angle=0]{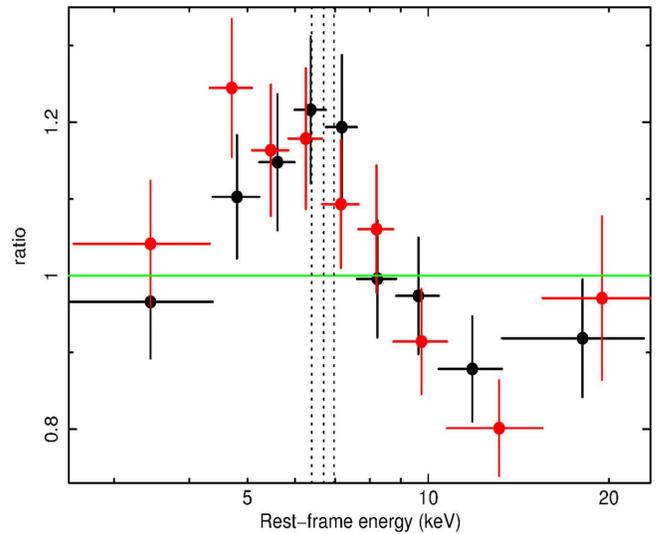}
   \caption{Ratio between the \emph{NuSTAR} spectrum and an
     absorbed $\Gamma=2$ power-law continuum model. The \emph{NuSTAR}
     FPMA and FPMB data are shown in black and red, respectively. The vertical
     dotted lines indicate the rest-frame energies for Fe K$\alpha$ at E$=$6.4~keV, Fe~XXV
     He$\alpha$ at E$=$6.7~keV, and the Fe~XXVI Ly$\alpha$ at
     E$=$6.97~keV, respectively. The data are binned to a minimum signal-to-noise of 13 for clarity. The P-Cygni
     emission/absorption profile is evident in the data.}
    \end{figure}

The output parameters of the \emph{XSTAR} fit are the column density,
ionization parameter, and the observed absorber redshift $z_o$. 
The ionization parameter is defined as $\xi$$=$$L_\mathrm{ion}/(n r^2)$ erg~s$^{-1}$~cm
(Tarter, Tucker \& Salpeter 1969) where $n$ is the number density of the material, and $r$ is the distance of
the gas from the central source. The observed absorber redshift is related to the intrinsic absorber
redshift in the source rest frame $z_a$ as $(1 + z_o) = (1 + z_a)(1 + z_c)$, where
$z_c$ is the cosmological redshift of the source. The velocity can then
be determined using the relativistic Doppler formula $1 + z_a = (1 - \beta/1 + \beta)^{1/2}$, where $\beta = v/c$. 



We obtain a high fit improvement of $\chi^2/\nu = 189/180$ ($\Delta
\chi^2 / \Delta \nu = 21/3$) indicating a
requirement of the \emph{XSTAR} absorber at the 99.97\% according to
the F-test ($\simeq 4 \sigma$). The best fit
parameters are a column density of $N_H = (2.0^{+0.5}_{-0.3})\times
10^{24}$~cm$^{-2}$, an ionization parameter of
log$\xi$$=$$5.3^{+0.7}_{-0.5}$~erg~s$^{-1}$~cm, and an observed absorber
redshift of $z_o = -0.18\pm0.03$. The power-law continuum slope is
$\Gamma = 2.15\pm0.04$.

We then calculated an \emph{XSTAR} emission table with the same
parameters as for the absorption table and include it in the model to
characterize the possible emission from the wind. The parameters of the
emission table are column density, ionization, redshift, and
normalization. We approximate the P-Cygni profile linking the ionization
parameter and column density between the emission and absorption
tables. We fix the redshift of the emitter to the cosmological redshift of the
source and in order to parameterize the broadening of the emission
feature we convolve the emission table with a Gaussian broadening profile.

The inclusion of the emission table provides an additional fit
improvement of $\chi^2/\nu = 171/178$ ($\Delta \chi^2 / \Delta \nu =
18/2$), which corresponds to a high statistical requirement of 99.98\%
according to the F-test ($\simeq 4 \sigma$). The slope of the power-law continuum
is now well constrained to be $\Gamma = 2.0\pm0.1$. The best fit
column density and ionization parameters are $N_H = (3.2\pm1.5)\times
10^{24}$~cm$^{-2}$ and log$\xi$$=$$4.0^{+1.2}_{-0.3}$~erg~s$^{-1}$~cm,
respectively.

The observed redshift of the absorber is estimated to be $z_o =
-0.083^{+0.122}_{-0.058}$, which corresponds to an outflow velocity of
$v_{\text{out}} = 0.253^{+0.061}_{-0.118}$~c.
The width of the emission line is $\sigma_E = 1.5^{+0.6}_{-0.4}$~keV,
which corresponds to a velocity width of $\sigma_v =
0.21^{+0.09}_{-0.05}$~c.

The \emph{NuSTAR} spectrum and the best fit wind model including both emission and absorption is
shown in Fig.~3. The combined statistical requirement of both emission and
absorption components from the wind is very high, at the level of
$5\sigma$. We note that the best fit parameters derived with
\emph{NuSTAR} are fully consistent with those estimated from the
\emph{Suzaku} spectrum and reported in Tombesi et al.~(2015). 

We note that the slope of the X-ray power-law continuum estimated to be
$\Gamma = 2.0\pm0.1$ is consistent with the typical value for Seyferts
and qusars (e.g., Dadina 2008). IRAS~F11119$+$3257 is potentially
accreting close to its Eddington value (Tombesi et al.~2015). Other
very luminous quasars accreting at or beyond their Eddington value
have been recently observed to host powerfult UFOs and potentially
slightly steeper power-law slopes of $\Gamma \simeq 2.3$ when caught in high flux states
(e.g., Lanzuisi et al.~2016; Matzeu et al.~2017).

The normalization of the emission table is $N =
(7.2^{+2.5}_{-5.3})\times 10^{-5}$. The normalization of the photo-ionized emission component is defined
within the \emph{XSTAR} code as $N = f L_{ion}/D^2$, where $f =
\Omega/4\pi$ is the covering fraction of the material, $L_{ion}$ is
the ionizing luminosity in units $10^{38}$ erg~s$^{-1}$ from 1 to
1,000 Ryd (1 Ryd is 13.6~eV) and $D$ is the distance of the observer
to the source in kpc. This assumes a uniform, spherically symmetric,
Compton-thin shell of material, which may be somewhat an over-simplification
for the emission from a disk wind. However, given the limited energy
resolution of \emph{NuSTAR} it should still provide a reasonable
approximation of the emission line profile in a P-Cygni configuration.  

If let free, the column density of the emitter and the normalization are
degenerate. Assuming a column density linked to the one of the
absorber, as in the present case, we obtain an upper limit on
the covering fraction $f$. Instead, assuming a covering fraction of
unity, we would estimate a lower limit in
the column density.

Substituting the appropriate values of the angular diameter
distance $D \simeq 6.3 \times
10^5$ kpc and the extrapolated power-law ionizing luminosity of
$L_{ion} \simeq 7 \times 10^6 (\times 10^{38})$~erg~s$^{-1}$, we
obtain a normalization of $N\simeq 2\times 10^{-5}$ for
$f = 1$. Comparing this value with the previous estimate we derive the
covering fraction of the wind is consistent with unity within the uncertainties.
Alternatively, fixing the normalization to the value
derived for $f=1$ and performing a new fit, we estimate a lower limit on the column density of
the emitter of $N_H \ge 1.3\times 10^{24}$~cm$^{-2}$. Both fits provide
the same $\chi^2/\nu$ statistic. These tests support the fact that the
emitter has both a high covering fraction ($f \simeq 1$) and a high
column density ($N_H \ge 10^{24}$~cm$^{-2}$). 

Although the wind parameters are well approximated with \emph{XSTAR}
absorption and emission tables, we note that a more physical model of a high column density wind
should include both Compton scattering and line
emission/reflection. Compton scattering may introduce a continuum
break at an energy beyond the observed \emph{NuSTAR} bandpass at E$>$50~keV and it
may contribute to the broadening of the lines of less than
$\Delta E$$\simeq$0.3~keV. A few detailed models have been reported in
the literature, but they are not publicly available for use in XSPEC
and they require a relative fine-tuning of the parameters (e.g., Sim
et al.~2010; Hagino et al.~2015). Even though our main conclusion will
likely not change, as shown for instance for the wind in the quasar PDS~456 (e.g.,
Nardini et al.~2015; Hagino et al.~2015), we plan to employ more sophisticated wind models in future works.

\subsection[]{Ionized disk reflection}

We note that the curvature in the X-ray spectrum might be alternatively
modeled as due to strong gravitational reflection from the inner
accretion disk. Even though this possibility was already disfavored
from the time-resolved spectral analysis of the \emph{Suzaku} data
reported in Tombesi et al.~(2015), here we consider again this possibility using the average \emph{NuSTAR} spectrum. 

We use the most accurate relativistic reflection code available in the
literature, the \emph{relxilllp} model (Garc{\'{\i}}a et al.~2014). This model considers
a lamp-post geometry in which the compact X-ray emitting source is located on
the rotation axis of the black hole at a certain height in units of
gravitational radii $r_g = GM_{BH}/c^2$. The reflection fraction and
emissivity index are estimated depending on the source
height and black hole spin. In order to allow for the largest possible relativistic broadening, we consider the case of a maximally spinning black
hole. We also assume a typical outer disk radius of
$r_{\text{out}} = 400 r_g$, and the inner radius $r_{in}$ is linked to the
innermost stable circular orbit. We consider standard Solar
abundances, but we checked also the case of free Fe abundance. The
high-energy cutoff is assumed at the typical value of $E_c \simeq 100$~keV
(Malizia et al.~2014; Fabian et al.~2015; Ricci et al.~2017).
Then, the free parameters of the model are the height of the illuminating source $h$, the disk inclination angle $i$, the ionization parameter log$\xi$, and the
normalization.

Initially, we checked if the relativistically broadened disk reflection model alone could describe the \emph{NuSTAR}
spectrum, considering a Solar abundance for iron. From a statistical
point of view, we obtain a good fit
($\chi^2/\nu = 176/180$) with an inclination of $i = 59^{+6}_{-8}$~degrees,
ionization parameter log$\xi$$=$$3.3\pm0.2$~erg~s$^{-1}$~cm, and
power-law slope $\Gamma = 2.0\pm0.1$. However, the fit is rather
extreme considering the other physical parameters. In fact, the height
of the compact emitting source is pegged to the hard lower limit of $h = 3 r_g$, with a
90\% upper limit of $h < 35 r_g$. Moreover, the reflection fraction would be rather
high, $R \simeq 4$, requiring a reflection dominated
spectrum. Leaving the Fe abundance free to vary it provides only a marginal
improvement to the fit ($\Delta \chi^2/\Delta \nu = 3/1$), with the
value pegged to the hard upper limit of 10. Recent papers questioned
the physical reliability of such extreme parameters (e.g., Dov{\v c}iak \& Done 2016).

  \begin{figure}[t!]
  \centering
   \includegraphics[width=8.5cm,height=7cm,angle=0]{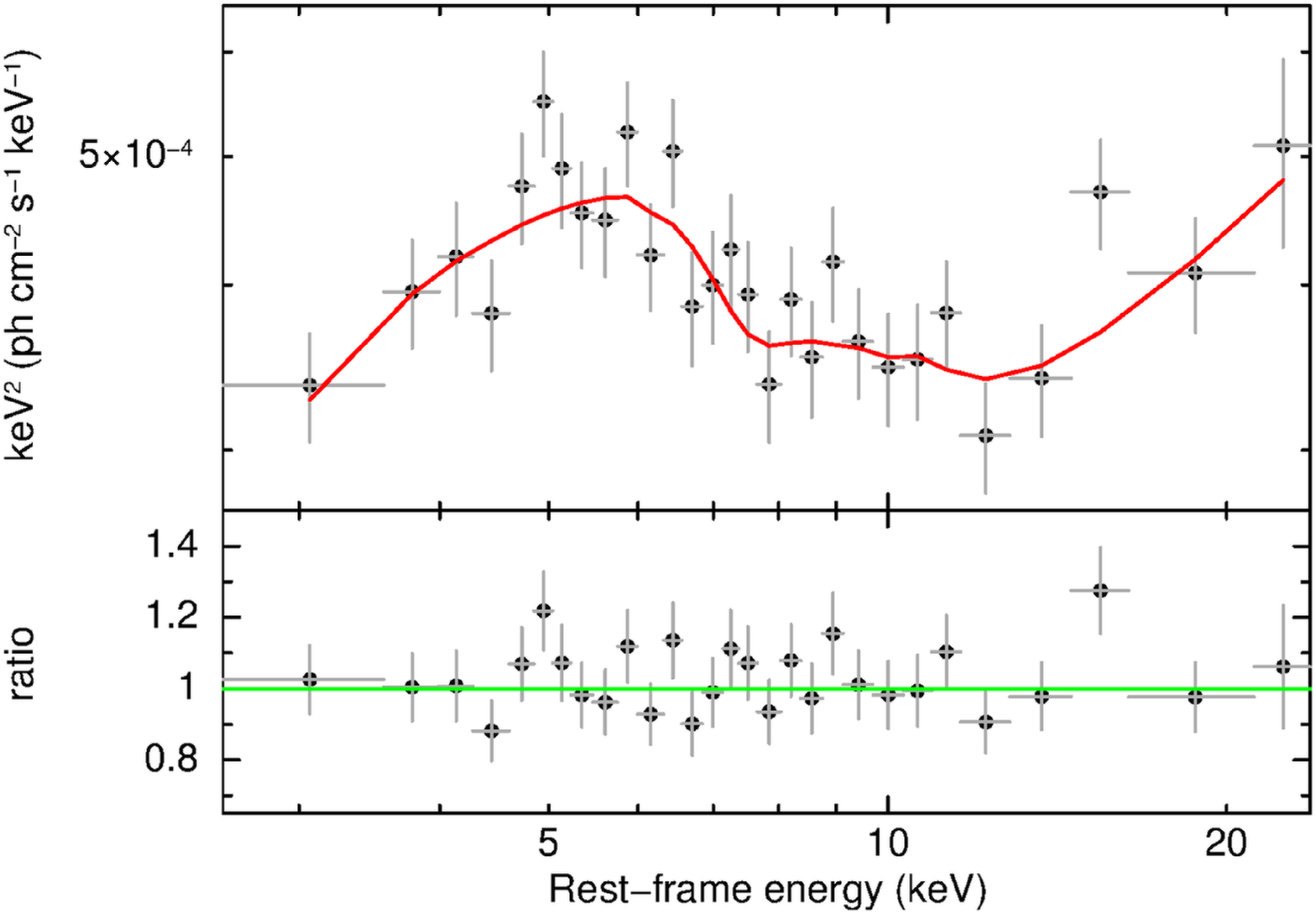}
   \caption{\emph{Upper panel:} combined \emph{NuSTAR} spectrum
     overlaid with the best fit ionized emission and absorption wind
     model. \emph{Lower panel:} ratio between the data and the
     best fit model. The data are binned to a
     minimum signal-to-noise of 10 for clarity.}
    \end{figure}

 We then checked the possibility to have both relativistic disk
 reflection and wind absorption combining both the \emph{relxilllp} emission
 and \emph{XSTAR} absorption tables. This is phneomologically similar to what we may expect to see from a
Compton-thick wind which is launched close to the black hole.
Indeed, we obtain a fit ($\chi^2/\nu =
 171/177$) with a disk inclination of $i = 60^{+27}_{-10}$~deg,
 ionization parameter log$\xi$$=$$3.5^{+0.6}_{-0.4}$~erg~s$^{-1}$~cm,
 and power-law slope $\Gamma = 2.0^{+0.5}_{-0.1}$. The source height
 is again pegged to the lower limit of $h = 3 r_g$, and we derive a 90\%
 upper limit of $h < 28 r_g$. The absorber is found to have a column
 density of $N_H = (1.7^{+2.2}_{-0.9})\times 10^{23}$~cm$^{-2}$,
 ionization parameter of
 log$\xi$$=$$2.5^{+0.9}_{-0.5}$~erg~s$^{-1}$~cm, and a less
 constrained velocity shift of $v_{out} =
 0.13^{+0.13}_{-0.11}$~c. 

From a purely statistical point of view, the relativistic disk reflection
plus wind absorption model is equivalent to the P-Cygni wind
emission/absorption model discussed in the previous section. 
This is similar to what we reported in Tombesi et al.~(2015) for the time-averaged
\emph{Suzaku} spectrum of IRAS~F11119$+$3257. However, in Tombesi et
al.~(2015) we demonstrated that the wind model was preferred to the
relativistic disk reflection model because it provided a simple explanation of the source flux variability as due to
a change in absorber column density. In contrast, within the relativistic reflection model, the resultant
variability pattern would be the opposite of that expected from
the relativistic light bending model (Miniutti \& Fabian 2004). Furthermore, the
extrapolated hard X-ray luminosity in the energy band
  E$=$14--195~keV from the fast wind model matches the value expected from mid-infrared emission line diagnostics, while
the relativistic reflection model underestimated this value by more
than one order of magnitude (Tombesi et al.~2015). 



\floattable 
\begin{deluxetable}{lccccc}[ht!]
\tablecaption{Average parameters of the ultra-fast outflow in
  IRAS~F11119$+$3257 observed with \emph{Suzaku} and \emph{NuSTAR}.}
\tablehead{
\colhead{Satellite} & \colhead{Year} & \colhead{log$\xi$} &
\colhead{$N_H$} & \colhead{$v_{out}$} & \colhead{$P$}\\
\colhead{} & \colhead{} & \colhead{(erg~s$^{-1}$~cm)} &
\colhead{($10^{24}$~cm$^{-2}$)} & \colhead{($c$)} & \colhead{($\sigma$)}
}
\startdata
\emph{Suzaku} & 2013 & $4.11^{+0.09}_{-0.04}$ & $6.4^{+0.8}_{-1.3}$ &
$0.255\pm0.011$ & 6.5\\
\emph{NuSTAR} & 2015 & $4.0^{+1.2}_{-0.3}$ & $3.2\pm1.5$ &
$0.253^{+0.061}_{-0.118}$ & 5.0\\
\enddata
\tablecomments{Columns: satellite name; observation year; ionization
  parameter; column density; outflow velocity; detection significance.}
\end{deluxetable}

\section[]{Discussion}

\subsection[]{The ultra-fast outflow}

In the previous sections we reported the analysis of the \emph{NuSTAR}
spectrum of IRAS~F11119$+$3257. The main objective of this project is
to perform an independent check of the results derived from the \emph{Suzaku} observation reported in
Tombesi et al.~(2015). Indeed, we clearly detect a highly ionized Fe K 
wind in the \emph{NuSTAR} spectrum. The best fit parameters are a
column density of $N_H = (3.2\pm1.5)\times 10^{24}$~cm$^{-2}$, an
ionization parameter of log$\xi$$=$$4.0^{+1.2}_{-0.3}$~erg~s$^{-1}$~cm, and an outflow
velocity of $v_{\text{out}} = 0.253^{+0.061}_{-0.118}$~c.

These wind parameters are consistent with the ones reported in Tombesi
et al.~(2015) for the average \emph{Suzaku} spectrum obtained in 2013,
see Table~1. The ionization and velocity of the wind seems to be stable on a
time-scale of at least two years, spanning between the \emph{Suzaku}
and \emph{NuSTAR} observations. Instead, already from the long \emph{Suzaku} observation, the absorber column
density was found to vary between $N_H \simeq (3-9)\times 10^{24}$
cm$^{-2}$ on a time-scale as short as three days. This may suggest
clumpiness in the wind, as already reported for other quasar disk
winds (e.g., Reeves et al.~2016; Matzeu et al.~2016).  

Given the best fit parameters of the wind estimated from \emph{NuSTAR}, we can estimate
the energetics of the wind following the approach described in Tombesi
et al.~(2015). The mass of the central super-massive black hole in
IRAS~F11119$+$3257 is estimated to be log$M_{BH} = 7.2\pm0.5$~$M_{\odot}$ (Kawakatu et al.~2007). A lower limit on the location of the absorber can be derived from the radius at which the observed
velocity corresponds to the escape velocity, $r = 2GM_{BH}/v_{out}^{2}
\simeq 7.4\times 10^{13}$~cm. Converting in units of Schwarzschild
radii, the wind launching radius is likely at a distance of $r \ge 16$~$r_s$ from the central black hole.

The mass outflow rate of the wind can be estimated considering the
equation $\dot{M}_{out} = 4\pi \mu m_p r N_H C_F v_{out}$, where $\mu =
  1.4$ is the mean atomic mass per proton, $m_p$ is the proton mass, and $C_F$ is
the wind covering fraction (Crenshaw \& Kraemer 2012). In spherical
symmetry this latter value corresponds to the solid angle subtended by
the wind of $\Omega = 4\pi C_F$. A high covering fraction $C_F \simeq
1$ was estimated in IRAS~F11119$+$3257 by Tombesi et al.~(2015) and in
the quasar PDS~456 (Nardini et al.~2015). Conservatively, here we
assume $C_F \simeq 0.5$ estimated from the fraction of sources with
detected ultra-fast outflows (UFOs) and warm absorbers (e.g., Tombesi et
al.~2010, 2013, 2014; Crenshaw \& Kraemer 2012; Gofford et al.~2013).
Therefore, we conservatively estimate a lower limit on the mass
outflow rate of $\dot{M}_{out} \simeq 0.5$~$M_{\odot}$~yr$^{-1}$. 

Given the high luminosity of this AGN and the SMBH mass estimate, this
source may likely be accreting at a rate close or slightly higher
thant Eddington. Mildly super-Eddington sources are likely to host
powerful disk winds and it is likely that the AGN radiation contributes
significantly to driving the wind (e.g., Matzeu et al.~2017). For a highly ionized wind driven mostly by radiation pressure due to
Compton scattering we have the relation $\dot{P}_{out} = \tau C_F
\dot{P}_{AGN}$ (e.g., Reynolds 2012; Tombesi et al.~2013, 2015; Matzeu
et al.~2017), where $\dot{P}_{out}$ and $\dot{P}_{AGN}$ are the
momentum rates of the wind and AGN radiation, respectively. Considering a covering fraction $C_F \simeq 1$ and the
optical depth of $\tau \simeq 2$ derived from the wind column density,
we can estimate $\dot{P}_{out} \simeq 2 \dot{P}_{AGN}$ to be the upper
limit for a radiation driven wind. Substituting $\dot{P}_{AGN} =
L_{AGN}/c$, $\dot{P}_{out} = \dot{M}_{out} v_{out}$, and $L_{AGN} \simeq 1.5\times 10^{46}$~erg~s$^{-1}$
(Veilleux et al.~2013), we can estimate an upper limit on the mass
outflow rate of $\dot{M}_{out} \simeq
2$~$M_{\odot}$~yr$^{-1}$. Therefore, the wind mass outflow rate is
likely in the range $\dot{M}_{out}$$\simeq$0.5--2~$M_{\odot}$~yr$^{-1}$.
The momentum flux (or force) and mechanical (or kinetic) power of the
wind are then estimated to be $\dot{P}_{out} \simeq
(2.5-10)\times 10^{35}$~dyne and $\dot{E}_{out}  = (1/2)\dot{M}_{out} v_{out}^2 \simeq (1-4)\times
10^{45}$~erg~s$^{-1}$, respectively. 

Comparing the wind energetics
with the AGN luminosity we obtain
$\dot{P}_{out} \simeq$0.5--2 $\dot{P}_{AGN}$ and
$\dot{E}_{out} \simeq$7--27\% $L_{AGN}$, respectively.
We note that the wind parameters estimated from \emph{NuSTAR} are
consistent with the ones derived from \emph{Suzaku} in Tombesi et
al.~(2015). In fact, in the \emph{Suzaku} case we obtained
$\dot{M}_{out} \simeq$1.5--4.5 $M_{\odot}$~yr$^{-1}$, $\dot{P}_{out}
\simeq$0.4--3 $\dot{P}_{AGN}$, and $\dot{E}_{out} \simeq$5--42\%
$L_{AGN}$, respectively. The UFO is consistent with
having a momentum rate equivalent to that of the AGN radiation and the
energetics is high enough to have an influence in AGN feedback (e.g.,
Di Matteo et al.~2005; Hopkins \& Elvis 2010; Gaspari et al.~2011).

\subsection[]{Connection with galaxy-scale molecular outflows}

Comparing the energetics of the X-ray UFO
detected with \emph{Suzaku} and the
OH molecular outflow detected with \emph{Herschel} in
IRAS~F11119$+$3257 we found that the favored interpretation of the
connection between the two winds was provided by the energy-conserving
relation $\dot{P}_{out} \simeq f(v_X/v_G) (L_{AGN}/c)$, where $v_X$ is the
velocity of the inner X-ray wind, $v_G$ is the velocity of the
galaxy-scale molecular outflow, and $f$ is the ratio between the
covering fractions of the outer molecular outflow and inner disk wind
(Tombesi et al.~2015). Instead, in the case of momentum-conserving, we
would expect a relation $\dot{P}_{out} \simeq L_{AGN}/c$.  

Here, we consider the wind parameters derived from \emph{NuSTAR} and
compare them with the latest estimates of the OH and CO molecular outflows in IRAS~F11119$+$3257 reported by
Veilleux et al.~(2017). The OH molecular outflow detected with
\emph{Herschel} is estimated to have a maximum velocity of
$v_{out}$$\simeq$1,000~km~s$^{-1}$, a mass outflow rate of $\dot{M}_{out}
\simeq$250--2,000 $M_{\odot}$~yr$^{-1}$, a momentum rate
$\dot{P}_{out} \simeq$3.5--25 $L_{AGN}/c$, a mechanical energy
$\dot{E}_{out}\simeq$0.5-5\% $L_{AGN}$, and a distance of $d$$\simeq$0.1--1~kpc from the
central black hole. The parameters of the CO molecular outflow
detected with \emph{ALMA} are a maximum velocity of
$v_{out}$$\simeq$1,000~km~s$^{-1}$, a mass outflow rate of $\dot{M}_{out}
\simeq$400--1,000 $M_{\odot}$~yr$^{-1}$, a momentum rate
$\dot{P}_{out} \simeq$8--16 $L_{AGN}/c$, a mechanical energy
$\dot{E}_{out}\simeq$0.8-2\% $L_{AGN}$, and a distance of $d$$\sim$4--15~kpc.

We note that in both cases the outflow momentum rate is about a factor
of ten higher than that in the radiation field and the energetics is
up to a few per-cent of the AGN luminosity. Considering the
energy-conserving relation previously described and substituting the relative parameters we find that
the inner UFO would indeed be able to drive the observed molecular
outflows with momentum rates $\dot{P}_{out}$$\simeq$15
$L_{AGN}/c$. 

However, we should bear in mind that the estimates of the mass outflow
rates may be affected by systematics and the conclusion may be less
stringent. In fact, the most conservative estimates derived by
Veilleux et al.~(2017) indicate that the OH and CO molecular outflows
may have momentum rates in the range $\dot{P}_{out}\simeq$1--6
$L_{AGN}/c$ and $\dot{P}_{out}\simeq$1.5--3 $L_{AGN}/c$,
respectively. Therefore, even if unlikely, the momentum-conserving
regime can not be fully excluded in this case. Moreover, we should
consider the caveat that the fast X-ray wind is observed now, while
the large-scale molecular outflows are probably an integrated effect of such winds over a much longer period of
time. Deep and spatially resolved soft X-ray and radio observations may help to
distinguish between energy- or momentum-conserving flows given that
the emission from the shocked gas is expected to be different in these
two cases (e.g., Bourne \& Nayakshin 2013; Nims et al. 2015). Some
attempts have been reported in the literature (e.g., Zakamska \&
Greene 2014; Tombesi et al.~2017).

\section[]{Conclusions}

In Tombesi et al.~(2015) we reported the detection of a powerful UFO in the \emph{Suzaku} X-ray
spectrum of the ultra-luminous infrared galaxy IRAS~F11119$+$3257. 
The comparison with a galaxy-scale OH molecular outflow observed with
\emph{Herschel} in the same source supported the energy-conserving
scenario for AGN feedback. In this work we perform an independent
check of the \emph{Suzaku} results using the higher
sensitivity and wider X-ray continuum coverage of \emph{NuSTAR}.

We clearly detect a highly ionized Fe K UFO in the 100~ks \emph{NuSTAR}
spectrum with parameters $N_H = (3.2\pm1.5)\times 10^{24}$~cm$^{-2}$,
log$\xi$$=$$4.0^{+1.2}_{-0.3}$~erg~s$^{-1}$~cm, and $v_{\text{out}} =
0.253^{+0.061}_{-0.118}$~c. The wind launching radius is likely at a
distance of $r \ge 16$~$r_s$ from the central black hole. The mass outflow rate is in the range
$\dot{M}_{out}$$\simeq$0.5--2~$M_{\odot}$~yr$^{-1}$. The UFO momentum
rate and power are $\dot{P}_{out} \simeq$0.5--2 $L_{AGN}/c$ and
$\dot{E}_{out} \simeq$7--27\% $L_{AGN}$, respectively. The UFO
parameters are consistent between the 2013 \emph{Suzaku} and the 2015
\emph{NuSTAR} observations. Only the column density is found to be
variable, possibly suggesting clumps in the wind. 


New multi-wavelength campaigns involving X-ray and optical, infrared,
millimeter and radio observatories of this source and a larger
population of luminous quasars will allow to constrain the physical details
of the AGN feedback processes and to determine if it is indeed widespread as
expected from numerical simulations of galaxy evolution (e.g., Fiore
et al.~2017). In the X-ray domain, transformative new results are expected from the unprecedented
spectroscopic capabilities of the planned X-ray astronomy recovery
mission (\emph{XARM}) (Hitomi Collaboration et al.~2016) and the
\emph{Athena} X-ray observatory (Nandra et al.~2013; Cappi et al.~2013).

\acknowledgments

F.T. and S.V. acknowledge support by the National Aeronautics and Space
Administration (NASA) through the NuSTAR award number
NNX15AV21G. F.T. acknowledges support by the Programma per Giovani Ricercatori - anno 2014 ``Rita Levi Montalcini''.

\end{document}